\documentclass[onecolumn,pra,aps,superscriptaddress,floatfix,nofootinbib,10pt]{revtex4}

\usepackage{amsfonts, amsmath, amssymb, amsthm}
\usepackage[mathscr]{euscript}
\usepackage{graphicx}

\usepackage{color}
\usepackage[normalem]{ulem}
\usepackage[toc,page]{appendix}
\usepackage[ocgcolorlinks,colorlinks=true,linkcolor=blue,citecolor=red]{hyperref}
\usepackage{latexsym}
\usepackage{natbib}
\usepackage{verbatim}
\usepackage{psfrag}

\newcommand{\be}{\begin{eqnarray}}
\newcommand{\ee}{\end{eqnarray}}
\newcommand{\unit}{1\!\!1}
\newcommand*{\Tr}[1]{\mathop{}\!\mathrm{Tr}{\left(#1\right)}}
\newcommand*{\sh}[1]{\mathop{}\!\mathrm{sinh}{\left(#1\right)}}

\newtheorem{llemma}{Lemma}

\newcommand*\dd{\mathop{}\!\mathrm{d}}

\newcommand{\braket}[2]{\langle #1  |#2\rangle}
\newcommand{\ketbra}[2]{|#1\rangle\langle #2|  }
\newcommand{\sandwich}[3]{\langle #1|#2 |#3\rangle  }

\newcommand{\mean}[1]{\ensuremath{\langle#1\rangle}}

\def\bra#1{\left<#1\right|}
\def\ket#1{\left|#1\right>}

\def\text#1{\textrm{#1}}

\begin{document}
\title{Robust Macroscopic Quantum Measurements in the presence of limited control and knowledge}

\author{Marc-Olivier Renou}
\affiliation{Groupe de Physique Appliqu\'ee, Universit\'e de Gen\`eve, CH-1211 Gen\`eve, Switzerland}
\author{Nicolas Gisin} 
\affiliation{Groupe de Physique Appliqu\'ee, Universit\'e de Gen\`eve, CH-1211 Gen\`eve, Switzerland}
\author{Florian Fr\"owis }
\affiliation{Groupe de Physique Appliqu\'ee, Universit\'e de Gen\`eve, CH-1211 Gen\`eve, Switzerland}

\date{\today}

\begin{abstract}
Quantum measurements have intrinsic properties which seem incompatible with our everyday-life macroscopic measurements.
Macroscopic Quantum Measurement (MQM) is a concept that aims at bridging the gap between well understood microscopic quantum measurements and macroscopic classical measurements.
In this paper, we focus on the task of the polarization direction estimation of a system of $N$ spins $1/2$ particles and investigate the model some of us proposed in Barnea et al., 2017.
This model is based on a von Neumann pointer measurement, where each spin component of the system is coupled to one of the three spatial components direction of a pointer. It shows traits of a classical measurement for an intermediate coupling strength.
We investigate relaxations of the assumptions on the initial knowledge about the state and on the control over the MQM.
We show that the model is robust with regard to these relaxations.
It performs well for thermal states and a lack of knowledge about the size of the system.
Furthermore, a lack of control on the MQM can be compensated by repeated "ultra-weak" measurements.
\end{abstract}

\maketitle

\section{Introduction}

In our macroscopic world, we constantly measure our environment. For instance, to find north with a compass, we perform a direction measurement by looking at the pointer. Yet, finding a quantum model for this kind of macroscopic measurement faces several problems. Many characteristics of quantum measurements seem to be incompatible with our intuitive notion of macroscopic measurements. For example, to perfectly measure two noncommuting observables is impossible in quantum mechanics and any informative measurement has a nonvanishing invasiveness. Thus, if it exists, such a model can not be of the standard projective kind. Although we have a good intuition of what such a measurement is, the natural characteristics it should satisfy are not obvious. Even if these characteristics can be rigorously formulated, it is not clear whether there exists a quantum model that satisfies them all.

For concreteness, quantum models for macroscopic measurements can be considered as a parameter estimation task. In this paper, we focus on the estimation of the direction of polarization of $N$ qubits, oriented in direction which is uniformaly choosen at random. The question of the optimal way to estimate $N$ qubit polarization is already well studied \cite{Massar95, Gisin99} and can be seen as part of a larger class of covariant estimation problem \cite{Chiribella05}. It is linked to covariant cloning \cite{Scarani05} and purification of state \cite{Cirac99}. In the limit of macroscopic systems, those optimal measurements are arbitrarly precise and potentially with low disturbance of the system \cite{Bagan05, Bagan06}. A tradeoff between the quality of the guess and the disturbance of the state has been demonstrated \cite{Sacchi07}, as well as an improvement of the guess when abstention is allowed \cite{Gendra13}. However, these optimal measurements may not be satisfying models of our everyday-life macroscopic measurements as it is not clear  how these optimal measurements could be physically implemented in a natural way. A first attempt to solve this issue has been to look for a reduction of the optimal POVM, which is continuous, into a POVM with a finite (and small) number of elements \cite{Latorre98, Chiribella07}. However, even if this reduction exists, the resulting POVM is difficult to interpret physically and to our best knowledge no family of reduced POVM for every $N$ exists.

In \cite{Barnea17}, we argue that a good model of a macroscopic measurement should be highly non invasive, collect a large amount of information in a single shot and be described by a "fairly simple" coupling  between  system  and  observer. Measurements which fulfills this requirements are called "Macroscopic Quantum Measurements" (MQM). 
Invasiveness seems to be difficult to satisfy with a quantum model. Indeed, the disturbance induced on the state by a measurement is generic in quantum mechanics. This has no counterpart in classical physics, where any measurement can ideally be done without disturbance of the system. However, it is now well known that this issue can be solved by accepting quantum measurements of finite accuracy. In \cite{Poulin05}, Poulin shows the existence of a trade-off between state disturbance and measurement resolution as a function of the size of the ensemble. One macroscopic observable can behave "classically", provided we measure it with sufficiently low resolution. Yet, the question is still open for several non commuting observables. Quantum physics allows precise measurements of only one observable among two non commuting ones. 

In this paper, we study the behavior of an MQM model for the measurement of the polarization of a large ensemble of $N$ parallel spin $1/2$ particles, which implies the measurement of the noncommuting spin operators. In this model, the measured system is first coupled to a measurement apparatus through an intuitive Hamiltonian already introduced in \cite{Dariano02}. Then, the apparatus is measured. 
We extend our previous study to more general cases. In \cite{Barnea17}, it was shown that this model allows good direction estimation and low disturbance for systems of $N$ parallel spin $1/2$ particles. This system can be interpreted as the ground state of a product Hamiltonian. Here, we generalize the scenario to thermal states.
We also study a different measurement procedure based on repeated weak measurements. 

The paper is structured as follows:
We first present a simplified technical framework which describes the measurement of a random direction for a given quantum state and observable. Considering an input state and an observable independent of the particle number and with no preferred direction, we show that the problem reduces to many sub-problems which correspond to systems of fixed total spin $j$.
Then, we quantitatively treat the case of the thermal state, which generalizes the $N$ parallel spin 1/2 particle for non-zero temperature, showing that the discussed MQM is still close to the optimal measurement.
In the proposed MQM, the precision of the estimated direction highly depends on the optimized coupling strength of the model. In section 4, we follow the ideas of \cite{Poulin05} and we show that one may relax this requirement by doing repeated "ultra-weak" measurements and a naive guess. We conclude and summarize in the last section.

\section{Estimation of a direction}

In this paper, we aim to study the behavior of a specific MQM model for a direction estimation task, e.g. the estimation of the direction of a magnet or a collection of spins. Hence, we first introduce an explicit (and specific) direction estimation problem, which is presented as a game. It concerns the direction estimation of a qubit ensemble. In the following, $S_{\vec{u}}=\vec{S}\cdot\vec{u}$ represents the spin operator projected in direction $\vec{u}$, i.e. the elementary generator of rotations around $\vec{u}$. 
For a given state $\rho_{\vec{u}}$ of $N=2J$ qubits, we say that $\rho_{\vec{u}}$ points in the direction $\vec{u}$ if it is positively polarized in the $\vec{u}$ direction, i.e. if $[\rho_{\vec{u}},S_{\vec{u}}]=0$ and $\Tr{\rho_{\vec{u}}S_{\vec{u}}}>0$. We consider the problem of polarization direction estimation from states which are all the same, but point in a direction which is choosen uniformally at random. This problem has already been widely studied \cite{Massar95}, \cite{Gisin99}, \cite{Chiribella05}, \cite{Bagan05}, \cite{Holevobook82} . We give here a unified framework adapted to our task.

\subsection{General framework}\label{generalframework}

We consider a game with a referee, Alice, and a player, Bob.  
Alice and Bob agree on some initial state $\rho_z$. In each round of the game, Alice chooses a direction $\vec{u}$ from a uniform distribution on the unit sphere. She rotates $\rho_z$ to $\rho_{\vec{u}}=\mathcal{R}_{\vec{u}}^\dagger \rho_z \mathcal{R}_{\vec{u}}$, where $\mathcal{R}_{\vec{u}}$ is a rotation operator which maps $\vec{z}$ to $\vec{u}$. She sends $\rho_{\vec{u}}$ to Bob, who measures it with some given measurement device characterized by a Positive Operator Valued Measure (POVM) $\Omega_r$. He obtains a result $r$ with probability $p(r|\vec{u})=\Tr{\Omega_r \rho_{\vec{u}}}$, from which he deduces $\vec{v}_r$, his guess for $\vec{u}$. Bob's score is computed according to some predefined score function $g(\vec{u},\vec{v}_r)=\vec{u}.\vec{v}_r$.
Given his measurement result, Bob's goal is to find the optimal estimate, i.e. the one which optimizes his mean score \footnote{Often the considered score is $F=\int \dd r\int \dd\vec{u}~ p(r|\vec{u})f(\vec{u},\vec{v}_r)$, where $f(\vec{u},\vec{v}_r)=|\braket{\vec{u}}{\vec{v}_r}|^2$ can be seen as the fidelity between qubits $\ket{\vec{u}}$ and $\ket{\vec{v}_r}$, where a unit vector is associated to the corresponding qubit via the Bloch sphere identification. As $F=\frac{1}{2}(1+G)$, this is equivalent. We chose this formulation for practical reason.}
\begin{equation}\label{score}
 G=\int \dd r\int \dd\vec{u}~ p(r|\vec{u})g(\vec{u},\vec{v}_r)
\end{equation}

For simplicity, we consider an equivalent but simplified POVM. In our description, Bob measures the system, obtain results $r$ and then post-process this information to find his guess $\vec{v}_r$. We now regroup all POVM elements corresponding to the same guess and label it by the guessed direction. Formally, we go from $\Omega_r$ to $O_{\vec{v}}=\int \dd r \Omega_r \delta(\vec{v}_r-\vec{v})$. 

Some assumptions are made about $\rho_z$ and $O_{\vec{v}}$.
We suppose that $\rho_z$ points in the $z$ direction. Moreover, we assume that $\rho_z$ is symmetric under exchange of particles, which implies $[\rho_z,S^2]=0$. Let $\ket{\alpha,j,m}$ be the basis in which $S_z$ and $S^2$ are diagonal (where $j\in\{J,J-1,...\}$ is the total spin, $\alpha$ the multiplicity due to particle exchange and $m$ the spin along $z$). Then $\rho_z$ is diagonal in this basis, with coefficients independent of $\alpha$, denoted as $c_m^j = \sandwich{\alpha,j,m}{\rho_z}{\alpha,j,m}$.

We also suppose that the measurement device does not favor any direction and treats each particle equally. Mathematically, it means that $O_{\vec{v}}$ is covariant with respect to particle exchange and rotations. Then, any POVM element is generated from one kernel $O_z$ and the rotations $\mathcal{R}_{\vec{v}}$: $O_{\vec{v}} = \mathcal{R}_{\vec{v}}^\dagger O_z \mathcal{R}_{\vec{v}}$ (for more technical details, see \cite{Holevobook82}).
With this, Eq.~(\ref{score}) simplifies to:

\begin{equation}\label{score2}
 G=\int \dd \vec{v}\int \dd\vec{u}~ p(\vec{v}|\vec{u})g(\vec{u},\vec{v}),
\end{equation}

\subsection{Score for given input state and measurement}

The following Lemma is already implicitly proven in \cite{Holevobook82}. 

\begin{llemma}\label{LlemmaG}
Bob's mean score is:
\begin{equation}\label{LemmaG}
G=\sum_{j}\frac{j A_j\Tr{\rho_z^{j}}}{j+1}\Tr{\frac{S_z}{j}\tilde{\rho}_z^{j}}\Tr{\frac{S_z}{j} \frac{O_z^{j}}{2j+1}}
\end{equation}

where $A_j={2J \choose J-j}-{2J \choose J-j-1}$ is the degeneracy of the multiplicity $\alpha$ in a subspace of given $(j, m)$, $O_z^{j}$ is the projections of $O_z$ over all subspaces of fixed $(\alpha, j)$, $\rho_z^{j}$ is the projection of $\rho_z$ over all subspaces of fixed $(\alpha, j)$ and $\tilde{\rho}_z^{j}=\frac{\rho_z^{j}}{\Tr{\rho_z^{j}}}$.
\end{llemma}

Lemma~\ref{LlemmaG} says that Bob cannot use any coherence between subspaces associated to different $(\alpha,j)$ to increase his score. In other words, the score Bob achieves is the weighted sum (where the weights are $\Tr{\rho_z^{\alpha,j}}$) of the scores $G^j$ Bob would achieve by playing with the states $\tilde{\rho}_z^{j}$. This property is a consequence of the assumption that no direction or particle is preferred by Bob's measurement or in the set of initial states. 
For self consistency, we prove this Lemma.

\begin{proof}
Bob's mean score is:
\begin{equation}\label{score3}
 G=\int \dd r\int \dd\vec{u}~ p(\vec{v}|\vec{u})g(\vec{u},\vec{v})=\int \dd v~ \Tr{O_{\vec{v}}\Gamma_{\vec{v}}},
\end{equation}
where $\Gamma_{\vec{v}}=\vec{v}\cdot\int \dd\vec{u}~\rho_{\vec{u}}~\vec{u}$. As $\rho_{\vec{u}}$ is the rotated $\rho_z$ and $O_{\vec{v}}$ is covariant, we have:
\begin{equation}
G=\Tr{O_z \Gamma_z}.
\end{equation}

Let $P_{\alpha,j}=\sum_m\ketbra{\alpha,j,m}{\alpha,j,m}$ be projectors, $\Gamma_{z}^{\alpha,j}=P_{\alpha,j}\Gamma_{z}P_{\alpha,j}$ and $O_{z}^{\alpha,j}=P_{\alpha,j}O_{z}P_{\alpha,j}$. Here, as $\rho$ and $O_z$ do not depend on the particle number, $\alpha$ is only a degeneracy.

As $\Gamma_z$ is invariant under rotation around $z$ and commutes with $S^2$, we have $\Gamma_z=\sum_{\alpha,j}\Gamma_z^{\alpha,j}$. Then $G=\sum_{\alpha,j}\Tr{O_{z}^{\alpha,j}\Gamma_{z}^{\alpha,j}}=\sum_{j} A_j \Tr{O_{z}^{j}\Gamma_{z}^{j}}$ where $O_{z}^{j},\Gamma_{z}^{j}$ are respectively the projections of $O_z$, $\Gamma_z$  over any spin coherent subspace of fixed $\alpha, j$.
Let $G^j=\Tr{O_{z}^{j}\Gamma_{z}^{j}}$.

$\Gamma_{z}^{j}=\sum_m c_m^j \int \dd\vec{u}~u_z \mathcal{R}_{\vec{u}}^\dagger \ketbra{\alpha,j,m}{\alpha,j,m} \mathcal{R}_{\vec{u}}$ is symmetric under rotations around $z$. Then, it is diagonal in the basis $\ket{\alpha,j,m}$ with fixed $j,\alpha$. As $\bra{\alpha,j,\mu}\int \dd\vec{u}~u_z \mathcal{R}_{\vec{u}}^\dagger \ketbra{\alpha,j,m}{\alpha,j,m} \mathcal{R}_{\vec{u}}\ket{\alpha,j,\mu}=\frac{m\mu}{j(j+1)(2j+1)}=\frac{m}{j(j+1)(2j+1)}\bra{\alpha,j,\mu}S_z^{\alpha,j}\ket{\alpha,j,\mu} $, we have:
\begin{equation}
\Gamma_{z}^{j}=\sum_{m} c_m^j \frac{m}{j(j+1)(2j+1)}S_z^{\alpha,j}
\end{equation}
and:
\begin{equation}
G^{j}=\frac{1}{j(j+1)(2j+1)}\Tr{S_z\rho_z^{j}}\Tr{S_z O_z^{j}}.
\end{equation}
\end{proof}

\subsection{State independent optimal measurement, optimal state for direction estimation}\label{State_indep_opt_measurement}

Given the state $\rho_z$, the measurement which optimizes Bob's score is the set of $\left\lbrace \Theta_{\vec{v}}^{\alpha,j} \right\rbrace$ such that $\Tr{S_z \Theta_z^{\alpha,j}}$ is maximal. The maximum is obtained when $\Theta_z^{\alpha,j}$ is proportional to a projector on the eigenspace of $S_z$ with the maximal eigenvalue, that is, for $\Theta_z^{\alpha,j}=(2j+1)\ketbra{\alpha,j,\pm j}{\alpha,j,\pm j}$. Here the sign depends of the sign of $\Tr{S_z\rho_z^{j}}$. In the following, we restrict ourselves to the case where the $\Tr{S_z\rho_z^{j}}$ are all positive (this is the case for the thermal state, considered below). Then:
\begin{equation}\label{Gopt}
G_{\mathrm{opt}}=\sum_{j}\frac{jA_j\Tr{\rho_z^{j}}}{j+1}\Tr{\frac{S_z}{j}\tilde{\rho}_z^{j}}.
\end{equation}
For $\rho_z=\ketbra{J,J}{J,J}$, the thermal state of temperature $T=0$, we find $G_{\mathrm{opt},T=0}=\frac{J}{J+1}$. Equivalently we recover the optimal fidelity $F_{\mathrm{opt},T=0}=\frac{1}{2}(1+G_{\mathrm{opt},T=0})=\frac{N+1}{N+2}$, already found in \cite{Massar95}. Asymptotically, we have $G_{\mathrm{opt},T=0}=1-1/J+O(1/J^2)$. This induces a natural characterization of the optimality of an estimation procedure. Writing $G_{T=0}$ as $G_{T=0}=1-\epsilon_J/J$ where $\epsilon_J= J(1-G_{T=0})\geq 1$, we say that the procedure is asymptotically optimal if $\epsilon_J=1+O(1/J)$ and almost optimal if $\epsilon_J-1$ is asymptotically not far from 0.

\subsection{Optimality of a state and a measurement for direction guessing}\label{Optimality}

Given the input state $\rho_z$, we can now compare the performances of a given measurement to the optimal measurement. 
From Eq.~(\ref{LemmaG}) and Eq.~(\ref{Gopt}), we have, for an arbitrary measurement:

\begin{equation}\label{DeltaG}
\Delta G\equiv G_{\text{opt}}-G=\sum_{j}\frac{j A_j \Tr{\rho_z^{j}}}{j+1}\Tr{\frac{S_z}{j}\tilde{\rho}_z^{j}}\Tr{\frac{S_z}{j} \frac{\Theta_z^j-O_z^{j}}{2j+1}}.
\end{equation}

For every $j$, the three terms of the product are positive. Then, qualitatively, the measurement is nearly optimal if for each $j$, the product of the three is small. We give here the interpretation of each of these terms:
\begin{itemize}
\item $A_j$ is the degeneracy under permutation of particles (labeled by $\alpha$) and $\Tr{\rho_z^{j}}$ the weight of $\rho_z$ over a subspace $j,\alpha$. Hence the first term, bounded by $j/(j+1)$, only contains the total weight of $\rho_z$ over a fixed total spin $j$. Hence, it is small whenever $\rho$ has little weight in the subspace $j$.

\item $\Tr{\frac{S_z}{j}\tilde{\rho}_z^{j}}$ is small whenever the component of $\rho_z$ on the subspace of total spin $j$, $\rho_z^{j}=P_z\rho_z P_z$, is small or not well polarized. It is bounded by 1. When $\rho_z^{j}$ is not well polarized, the optimality of the measurement in that subspace makes little difference. Then, this second term characterizes the quality of the component $\rho_z^{j}$ for the guess of the direction.

\item The last term is small when $O_z^{j}$ is nearly optimal and is also bounded by 1. More exactly, as $O_z^{j}$ is a covariant POVM, we have $\Tr{O_z^{j}}=2j+1$ and all diagonal coefficients are positive. Because of $S_z/j$, $O_z^{j}$ is (nearly) optimal when it projects (mainly) onto the subspace of $S_z$ with the highest eigenvalue. POVMs containing other projections are sub-optimal. This effect is amplified by the operator $S_z$: the further away these extra projections $\propto\ketbra{j,m}{j,m}$ are from the optimal projector $\propto\ketbra{j,j}{j,j}$ (in the sense of $j-m$), the stronger the sub-optimality is. Then, the last term corresponds to the optimality of the measurement component $O_z^{j}$ for the guess of the direction.
\end{itemize}
Interestingly, we see here that the state and measurement "decouple": The optimal measurement is independent of the considered state. However, if the measurement is not optimal only for subspaces where $\rho_z$ has low weight or is not strongly polarized, it will still result in a good mean score.

\subsection{Estimation from a thermal state}

We now consider the case where the game is played with a thermal state (with temperature $T=1/\beta$) of $N=2J$ spins:
\begin{equation}
\rho_z=\frac{1}{Z}\left(e^{-\beta \sigma_z/2}\right)^{\otimes N}=\frac{1}{Z}\sum_{\alpha,j,m}e^{-\beta m}\ketbra{\alpha,j,m}{\alpha,j,m},
\end{equation}
where $Z=\left(2 \text{cosh}(\beta/2)\right)^N$ is the partition sum.
$\rho_z$ is clearly invariant under rotations around $z$ and symmetric under particle exchange.
For later purpose, we define $f_j(\beta)=Z \Tr{\frac{S_z}{j}\rho_z^{\alpha,j}}=\Big[(1+j)\sh{j\beta}-j\sh{(1+j)\beta}\Big]/\Big(2j\sh{\beta/2}^2\Big)$.

Eq.~(\ref{LemmaG}) now reads 
\begin{equation}\label{GT=0}
G_{T=0}=\frac{J}{J+1}\Tr{\frac{S_z}{J}\frac{O_z}{2J+1}},
\end{equation}
and for any temperature $\beta$:
\begin{equation}
 G=\frac{1}{Z}\sum_{j}A_j~f_j(\beta)~G^j_{T=0},
\end{equation}
with the optimal measurement, $G_{\mathrm{opt},T}=\frac{1}{Z}\sum_j\frac{jA_j}{j+1}f_j(\beta)$.
Note that for low temperatures, this expression can be approximated with $\mean{J}_\beta/(\mean{J}_\beta+1)$, where $\mean{J}_\beta$ is the mean value of the total spin operator for a thermal state.

\section{A Macroscopic Quantum Measurement (MQM)}

\subsection{The model}
In the following, we consider a model already introduced in \cite{Dariano02, Barnea17} for polarization estimation. It is adapted from the Arthur Kelly model, which is designed to simultaneously measure momentum and position \cite{Arthurs65, Pal11, Levine89}. The model is expressed in the von Neumann measurement formalism \cite{Neumannbook55, Buschbook91, Peresbook02}. The measurement device consists of a quantum object -the pointer- which is first initialized in a well-known state and coupled to the system to be measured. At last, the pointer is measured in a projective way. The result of the measurement provides information about the state of the system. Tuning the initial state of the pointer and the strength of interaction, one can model a large range of measurements on the system, from projective measurements which are partially informative but destruct the state to weak measurements which acquire few information but do not perturb much.

More specifically, to measure the direction of $\rho_{\vec{u}}$, we use a pointer with three spatial degrees of freedom:
\begin{equation}
\ket{\phi}=\frac{1}{(2\pi\Delta^2)^{3/4}}\int \dd x\dd y\dd z e^{-\frac{x^2+y^2+z^2}{4\Delta^2}}\ket{x}\ket{y}\ket{z},
\end{equation}
where $x,y,z$ are the coordinates of the pointer. The parameter $\Delta$ in $\ket{\phi}$ represents the width of the pointer: A small $\Delta$ corresponds to a narrow pointer and implies a strong measurement, while a large $\Delta$ gives a large pointer and a weak measurement.
The interaction Hamiltonian reads:
\begin{equation}
H_{\text{int}}=\vec{S}\cdot\vec{p}\equiv p_x\otimes S_x + p_y\otimes S_y + p_z\otimes S_z,
\end{equation}
where $p_x,p_y,p_z$ are the conjugate variables of $x,y,z$. A longer interaction time or stronger coupling can always be renormalized by adjusting $\Delta$. Hence, we take the two equal to 1. 
Finally, a position measurement with outcome $\vec{r}$ is performed on  the pointer. 
The POVM elements associated to this measurement are $O_{\vec{r}}=E_{\vec{r}}E_{\vec{r}}^\dagger$, where the Krauss operator $E_{\vec{r}}$ reads:
\begin{equation}\label{defKrauss}
E_{\vec{r}}\propto\int \dd \vec{p}e^{i\vec{r}\cdot\vec{p}}e^{-\Delta^2p^2 e^{-i\vec{p}\cdot\vec{S}}}
\end{equation}
The POVM associated to this model is already covariant. Indeed, the index of each POVM element is the direction of guess
\footnote{To exactly obtain the form given in Sec.~\ref{generalframework}, one has to define introduce $O_{\vec{v}}=\int_{0}^{\infty}r^2 O_{\vec{r}}\dd r$, which is equivalent to identify each vector with its direction.}
 and any $O_{\vec{r}}$ is a rotation of $O_{z}$: $O_{\vec{r}}= \mathcal{R}_{\vec{r}}^\dagger O_{z} \mathcal{R}_{\vec{r}}$.

\subsection{Behavior for zero temperature states}\label{behavior_T0}

\begin{figure}
\center
\includegraphics[scale=.8]{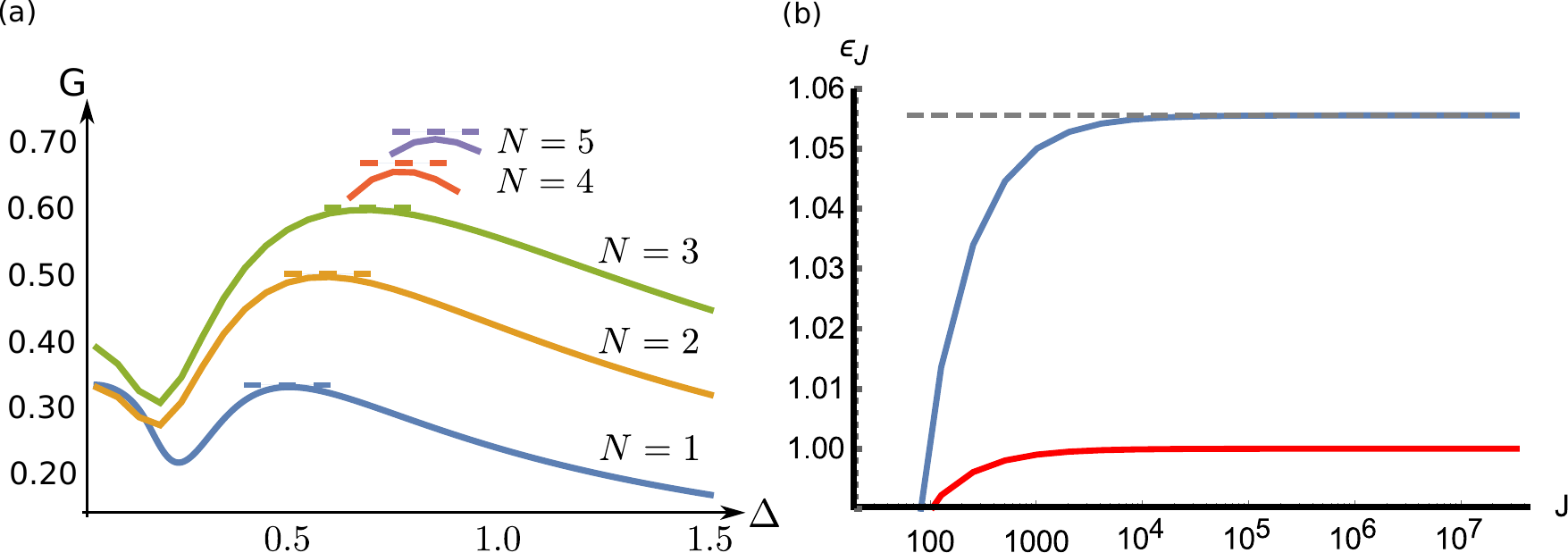}
\caption{(a) Mean score as a function of the pointer width $\Delta$ for various $N=2J$. The dashed lines correspond to the optimal value $G_\text{opt}$. (b) Scaling factor $\epsilon_J= J(1-G_J)$ from the approximate lower bound on the score $G$ (upper,  blue curve) compared to the optimal scaling factor $J(1-G_{\text{opt}})$(lower, red curve). For large $J$, $\epsilon_J$ seems to go to $19/18$ (dashed line). See Sec.~\ref{behavior_T0} for further details.}
\label{epsN_LargeJ_strange_behavior}
\end{figure}

At zero temperature, it is already known that the score obtained for a game where Bob does the MQM remains close to the optimal one. In our previous study \cite{Barnea17}, we demonstrated a counter-intuitive behavior of the quality of the guess:  a weaker coupling strength can achieve better results than a strong coupling, see Fig.~\ref{epsN_LargeJ_strange_behavior}(a). In particular, we show that for well chosen finite coupling strength, the score of the guess is almost optimal. The optimal value of the coupling is $\Delta=\sqrt{J/4}$: It scales with the square root of the number of particles.

Additional calculations confirm this first conclusion (see Fig.~\ref{epsN_LargeJ_strange_behavior}(b)). Exploiting the conclusion of the discussion of Sec.~\ref{Optimality}, we only considered the first diagonal coefficient of $O_z$, $o_J=\sandwich{J,J}{O_z}{J,J}$, to lower bound the performence of the POVM
\footnote{This method is equivalent to the one previously used in \cite{Barnea17}. There, $\bra{z}^{\otimes N}\Omega_{\vec{r}}\ket{z}^{\otimes N}$ is lower bounded by $|\bra{z}^{\otimes N}E_{\vec{r}}\ket{\vec{r}}^{\otimes N}|^2$, but as the Krauss operator is diagonal, this last term is nothing else than $o_J|\bra{z}^{\otimes N}\ket{\vec{r}}^{\otimes N}|^2$.}.
Numerical simulations suggest that for a coupling strength $\Delta=\sqrt{J/4}$, only considering the bound over $o_J$, $G_{T=0}$ develops as $G_{T=0}=1-\epsilon_J/J$ with $\epsilon_J= J(1-G_J)\lesssim 19/18$ for large $J$. Hence, the asymptotic difference between $G_{\mathrm{opt},T=0}$ and $G_{MQM,T=0}$ is such that $J\Delta G_{T=0}$ remains bounded, in the order of 0.05.

From Eq.~(\ref{LemmaG}) and the discussion about Eq.~(\ref{DeltaG}), we see that, to achieve optimality the first diagonal coefficient, $o_J$ must be maximal
\footnote{We can interpret this physically. We see from Sec.~\ref{State_indep_opt_measurement} that the best covarient measurement is obtained from $O_z\propto\ketbra{J,J}{J,J}$. Other covarient measurements can be obtained with $O_z\propto\ketbra{m,J}{m,J}$ for $0 \leq m<J$. The coefficients $o_m$ can be interpreted as how much each of these measurements is done. The term $\ketbra{m,J}{m,J}$ can also be thought as the physical system used to measure. When it is highly polarized ($m=J$), the measurement is efficient. But when the polarization is low, the information gain is weak. E.g., $m=0$, we clearly see that all POVM elements are $\propto\unit$.
}
, that is equal to $2J+1$. When it is not the case, as $\Tr{O_z}=2J+1$, the difference $(2J+1)-o_J=\Tr{O_z}-o_J=\sum_{m\neq J}o_m$ is distributed between the other diagonal coefficients $o_m=\sandwich{J,m}{O_z}{J,m}$, for $m\neq J$.
The score achieved by the measurement is given by Eq.~(\ref{GT=0}): 
\begin{equation}
G_{T=0}=\Tr{\frac{S_z}{J} \frac{O_z^{J}}{2J+1}}=\frac{J}{J+1}\sum_m \frac{m}{J}\frac{o_m}{2J+1}.
\end{equation}
Our bound only considers the coefficient $o_J$. However, a simple calculation shows that this is enough to deduce the strict suboptimality of the measurement.
Indeed, one can derive:
\begin{align*}
\epsilon_J&=J\left(1-\frac{J}{J+1}\left(\frac{o_J}{2J+1}+\sum_{m\neq J}\frac{m}{J}\frac{o_m}{2J+1}\right)\right)\\
&\geq J\left(1-\frac{J}{J+1}\left(\frac{o_J}{2J+1}+\frac{J-1}{J}\left(1-\frac{o_J}{2J+1}\right)\right)\right)\\
&\geq 2-\frac{o_J}{2J+1}+o(1),
\end{align*}
where $o(1)\rightarrow 0$ when $J\rightarrow \infty$.
Hence if $o_J$ is not asymptotically $2J+1$, $\epsilon_J$ cannot be asymptotically 1.

In the following, we show that a lower bound on G for thermal states can be calculated with methods based on the $T=0$ case.

\subsection{Behavior for finite temperature states}\label{behaviorTnon0}

As it is build from the spin operators only, the measurement scheme depends only on the properties of the system with respect to the spin operators.
More precisely, for a given system size $N=2J$, we consider the basis $\{\ket{\alpha,j,m}^{(N)}\}$ and  for given total spin $j$ and permutation multiplicity $\alpha$ the projector $P_{\alpha,j}^{(N)}=\sum_{\alpha,j}\ketbra{\alpha,j,m}{\alpha,j,m}^{(N)}$.
Then, the projection of Eq.~(\ref{defKrauss}) for $N=2J$ spins onto the subspace $j,\alpha$ is equivalent to the projected Krauss operator for $n=2j$ spins onto $j$:
\begin{equation}\label{independentN}
P_{\alpha,j}^{(N)}E_{\vec{r}}^{(N)}P_{\alpha,j}^{(N)}\equiv P^{(n)}_{j}E_{\vec{r}}^{(n)}P^{(n)}_{j},
\end{equation}
where the equivalence $\equiv$ is interpreted as $\ket{\alpha,j,m}^{(N)}\equiv\ket{m}^{(n)}$ (there is no multiplicity for $n$ and $j=n/2$).

\begin{figure} 
\center
\includegraphics[scale=0.4]{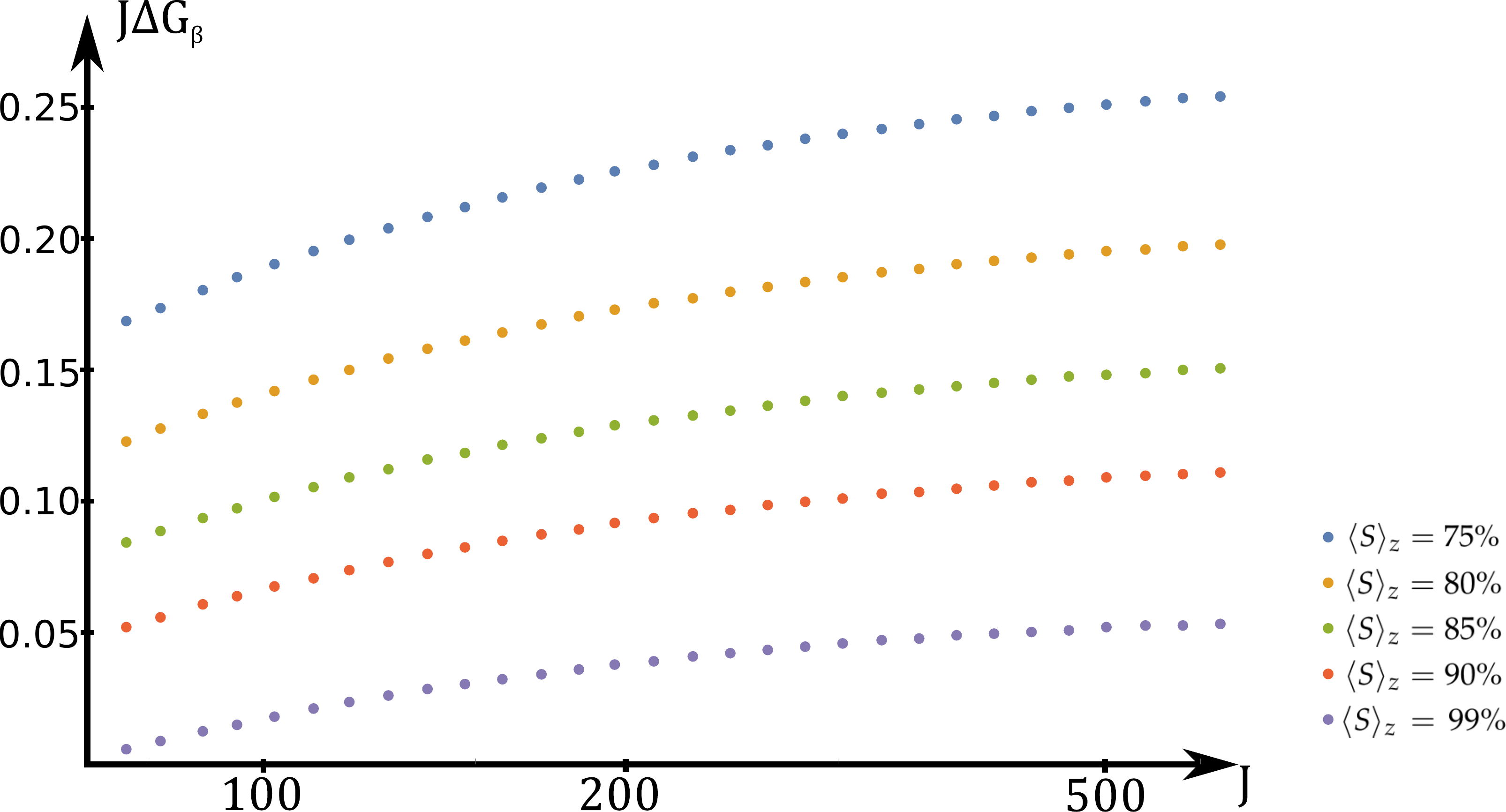}
\caption{$J \Delta G$ (Eq.~(\ref{DeltaG})) as a function of $J$, for various $\beta$ chosen such that $\left<S_z\right>=J \tanh{\beta/2}$. The MQM is close to optimal even for finite temperature. See Sec.~\ref{behaviorTnon0} for further details.}
\label{thermal}
\end{figure}
For non zero temperature, we adapt the numerical estimation model of \cite{Barnea17}. Due to Lemma~\ref{LlemmaG} and Eq.~(\ref{independentN}), we can directly exploit the same model and combine the results for the different subspaces for given $j$.
However, in this case, we are limitated by the choice of the coupling strength $\Delta$ of the pointer with the system. At zero temperature, only the total spin subspace which corresponds to $j=J$ is involved. The optimal coupling strength is then $\Delta=\sqrt{J/4}$. For a non zero temperature, all possible $j$ appears and the value of $\Delta$ cannot be optimized for each one.  Our strategy is to chose the optimal coupling value for the equivalent total spin $J_{\mathrm{eq}}$ satisfying $\mean{S^2}=J_{\mathrm{eq}}(J_{\mathrm{eq}}+1)$, which can be deduced from $\mean{S^2}=\frac{1}{1} (3J+J(2J-1)\text{tanh}^2{\beta/2})$ (for a thermal state). Depending on the sensitivity of the MQM guessing scheme with respect to a change in the value of $\Delta$, this method may work or not. Numeric simulations show that a change of order $O(\sqrt{J})$ perturb the score. However, one can hope that for smaller variation, the perturbation is insignificant.

We tested the method for different values temperature $T=1/\beta$ corresponding to spin polarization $\left<S_z\right>=J \tanh{\beta/2}$. 
We find again that the asymptotic difference between $G_{\mathrm{opt}}$ and $G_{MQM}$ is small. More precisely, Fig.~\ref{thermal} shows $J\Delta G_{\beta}$ as a function of $J$, for different temperature corresponding to $\mean{S_z}=c J$, for various $c$. 
For each $\Delta$, the error $J\Delta G_{\beta}$ seems to be bounded for large $J$. 

\section{Estimation of a direction through repeated weak measurements}

In the previous section, we considered a specific MQM and studied the mean score of the state direction for pure states as well as for more realistic thermal states. We compared it to its optimal value, obtained with the optimal theoretical measurement. We showed that the difference remained bounded. As the model makes use of a simple Hamiltonian coupling between system and observer, it satisfies the requirements of an MQM as stated in the introduction for thermal states.

However, this model requires that three one-dimensional (1D) pointers (or equivalently one three-dimensional (3D) pointer) are coupled to the system at the very same time, to be then measured. This requirement is difficult to meet. Moreover, an optimized coupling strength between system and pointer is necessary: the pointer width has to be $\Delta=\sqrt{J/4}$ within relatively tight limits. This requires a good knowledge about the system to be measured (its size, its temperature, ...) and fine control over the measurement. Following \cite{Poulin05}, we can overcome this problem by implementing many ultra-weak measurements.
To this end, we focus on a relaxation of the measurement procedure, where we consider repeated very weak measurements (with $\Delta\gg \sqrt{J/4}$) in successive orthogonal directions on the state, which is gradually disturbed by the measurements. This idea has already been implemented experimentaly \cite{Hacohen16}. The guessed state is obtained by averaging the results in each of the three directions. Note that this is not optimal, as the first measurements are more reliable than the last. However, we show in the following that this intuitive approach gives almost optimal results. For simplicity, we restrict ourselves to the case of a perfectly polarized state, or equivalently a thermal state at a zero temperature.
\subsection{The model}

We modify the game considered so far in the following way. Bob now uses a modified strategy, in which he successively repeats the same measurement potentially in different measurement basis. 
First, he weakly couples the state to a 1D Gaussian pointer through an interaction Hamiltonian in some direction $w$.
The pointer state is 
\begin{equation}
\ket{\phi}=\frac{1}{(2\pi\Delta^2)^{1/4}}\int \dd we^{-\frac{w^2}{4\Delta^2}}\ket{w},
\end{equation} 
and the Hamiltonian reads
\begin{equation}
H_w\propto p_w\otimes S_w,
\end{equation}
where $w\in \{x,y,z\}$. 
Then, Bob measures the pointer. The post measured state is used again for the next measurement and is disturbed in each round. 
We first analytically derive the case where Bob only measures in one direction ($w=z$). Then, we consider the case where Bob does $t$ measurements successively in each orthogonal direction $x, y, z$. He obtains results $x_1, y_1, z_1, x_2, y_2, ..., z_t$ and estimates the direction with the vector which coordinates are the average of the $x_i$, the $y_i$ and the $z_i$.

\subsection{Measurement in a single direction}\label{ultraweak1D}

We first study the 1D case.
First, note that the optimal strategy when the measurement operators $O_r$ are required to measure in a fixed direction $z$ (i.e. $[O_r,S_z]=0$) is to measure the operators $S^j_z$: As the $O_r$ commutes with $S_z$, they can be simulated with a measurement of $S^j_z$.
The optimum is to answer $\pm z$ depending on the sign of the result. The obtained score is then $G=\frac{J}{2J+1}$ for integer $J=N/2$ and $G=\frac{2J+1}{4(J+1)}$ otherwise.

In our model, we consider an interaction Hamiltonian $H_w$ taken in a constant direction $w=z$. The total number of measurements is $t$. 

\begin{figure}
\center
\includegraphics[scale=0.5]{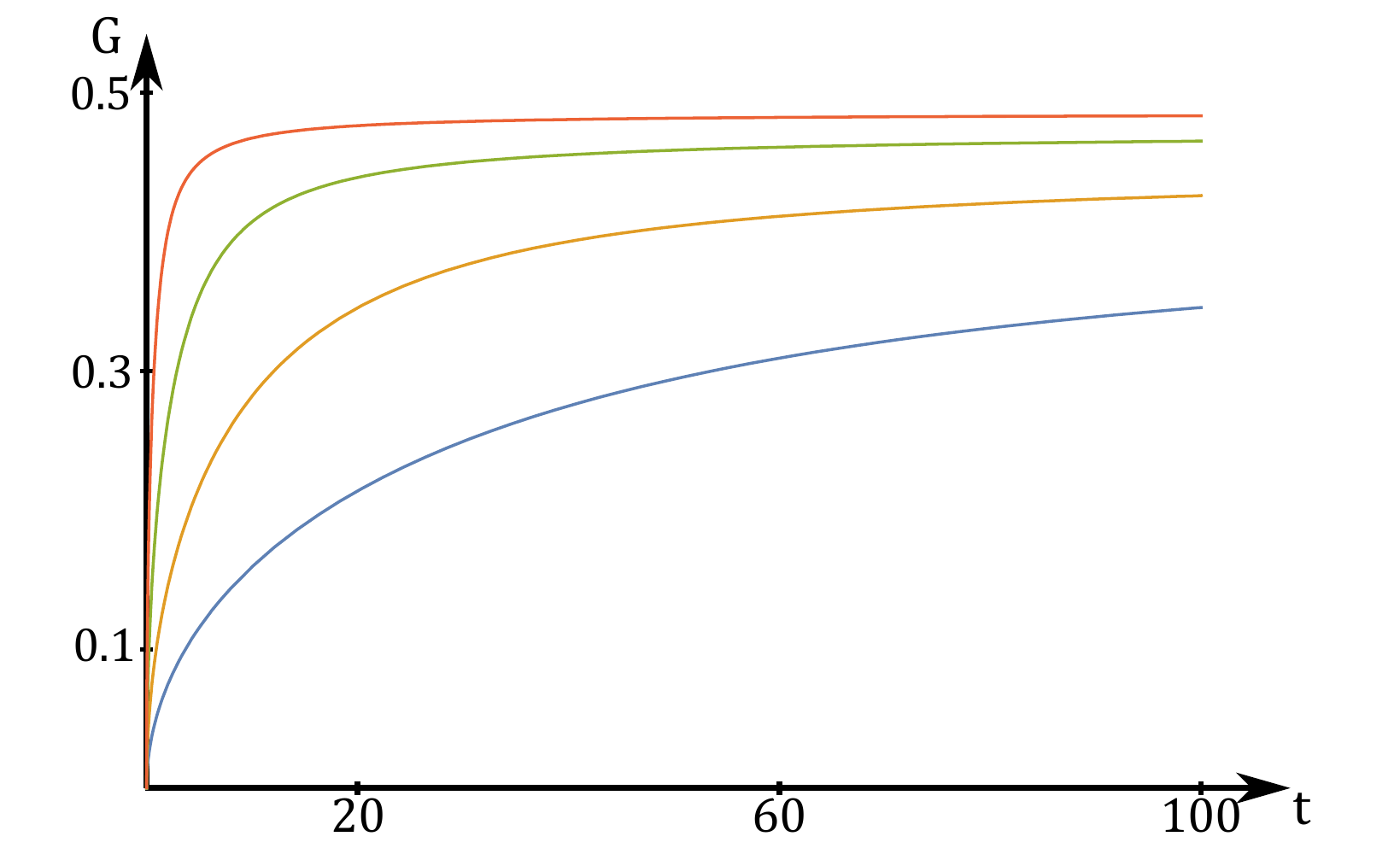}
\caption{Score for repeated weak measurement in a single fixed direction with $\Delta=10$ and $J=2, 4, 8, 16$. See Sec.~\ref{ultraweak1D} for further details.}
\label{repetitiveweak1d}
\end{figure}
 The measurement results form a vector $\vec{r}=\{r_1, ..., r_t\}$.  
The POVM of the full measure sequence is:
\begin{equation}
\Omega_{\vec{r}} =
  \begin{bmatrix}
    \ddots & & \\
    & F_m({\vec{r}}) & \\
    & & \ddots
  \end{bmatrix}
\end{equation}
where:
\begin{equation}
F_m(\vec{r})=\frac{1}{\left(\Delta\sqrt{2\pi}\right)^p}~ e^{\frac{-||\vec{r}-m\vec{1}||^2}{2\Delta^2}},
\end{equation}
where $\vec{1}=\{1, ..., 1\}$. As all measurements for each step commute, this case can be solved analytically. Note first that the ordering of the measurement results is irrelevant. From Eq.~(\ref{score}), we find:
\begin{align*}
G&=\frac{1}{(J+1)(2J+1)}\Tr{S_z O_z}\\
&=\frac{2}{(J+1)(2J+1)\left(\Delta\sqrt{2\pi}\right)^t}\int \dd \vec{r} \delta(\vec{v}_{\vec{r}}-\vec{z})e^{-\frac{||\vec{r}||^2}{2\Delta^2}}\left(\sum_{m>0} m~e^{\frac{-mt}{2\Delta^2}}\text{sinh}\left(\frac{m~\vec{r}\cdot\vec{1}}{\Delta^2}\right)\right),
\end{align*}
where $\vec{v}_{\vec{r}}$ is the optimal guess.
For $\vec{r}$ such that $\vec{r}\cdot\vec{1}\geq 0$, the optimal guess is clearly $\vec{v}_{\vec{r}}=\vec{z}$. By symmetry, $\vec{v}_{-\vec{r}}=-\vec{v}_{\vec{r}}$ and the optimal guess is $\vec{v}_{\vec{r}}=\text{sign}(\vec{r}\cdot\vec{1})\vec{z}$.
Then:
\begin{equation}
G=\frac{2}{(J+1)(2J+1)}\sum_{m>0}m~\text{erf}\left(\frac{m}{\Delta}\sqrt{\frac{t}{2}}\right)
\end{equation}
is easily computed by integration over $\vec{r}$ and by decomposition into its parallel and orthogonal components to $\vec{1}$.
We see here that the score only depends on the ratio $\frac{\sqrt{t}}{\Delta}$ and reaches the 1D strong measurement limit for $\frac{\sqrt{t}}{\Delta}\gg 1$ (see Fig.~\ref{repetitiveweak1d}). Here erf is the error function. We see that $G\rightarrow 1/2$ for $J\rightarrow \infty$, which is the optimal value for optimal measurements lying on one direction.

\begin{figure}
\center
\includegraphics[scale=0.4]{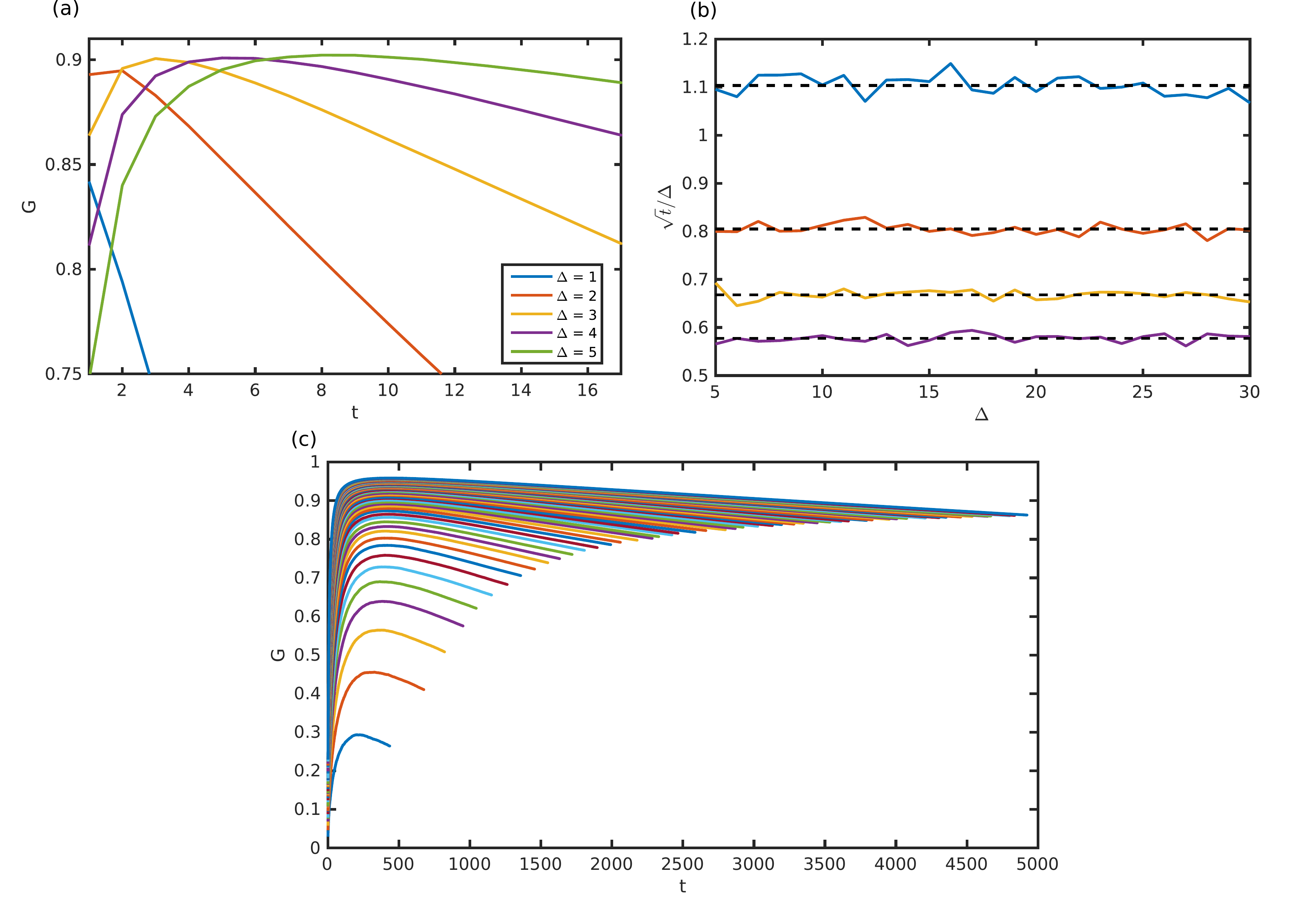}
\caption{(a) Score as a function of the number of measurements for $N=2J=20$. For each $\Delta$, there is an optimal repetition rate $t^\mathrm{max}$. The optimal score $G^{\mathrm {max}}$ saturates for $\Delta$ big enough. (b) Ratio $\sqrt{t^{\mathrm {max}}}/\Delta$ for $N=5, 10, 15, 20$. As in the 1D case, $\sqrt{t^{\mathrm {max}}}/\Delta$ is constant and only depends on $N$. (c) Score as a function of the number of measurements $t$, for $N=2J=1..50$ and $\Delta=8\sqrt{N}$. See Sec.~\ref{ultraweak3D} for further details.} 
\label{Goptsaturation}
\end{figure}

\subsection{Ultra-weak measurements in three orthogonal directions}\label{ultraweak3D}

We now study the relaxation of our initial MQM model. In this case, for large number of measurement $t$, we could not analytically derive the mean score. We hence implemented a numerical simulation of the model. 
We fix the number of qubits $N=2J$ and pointer width $\Delta$. The vector $\vec{u}$ is drawn at random on the Bloch sphere. Then, we simulate $\tau$ successive weak measurements in directions $x, y, z$ of the system  $\ket{\vec{u}}^{\otimes N}$. For each $t \leq \tau$, we guess $\vec{u}$ from the mean of the results for $x, y, z$ for measurements up to $t$.

For large $\Delta$, our procedure can be seen as successive weak measurements of the system. Each measurement acquires a small amount of information and weakly disturbs the state. We attribute the same weight to each measurement result to find the estimated polarization. As each measurement disturbs the state, this strategy is not optimal. However, keeping the heuristic of "intuitive measurement", we consider this guessing method as being natural.

The results from the numerical simulation suggest that for fixed number of particle $N=2J$ and fixed pointer width $\Delta$, the score as a function of $t$ increases and then decreases (see Fig.~\ref{Goptsaturation} (a)), which is intuitive. Indeed, for few measurements, the state is weakly disturbed and each measurement acquires only a small amount of information about the original state. Then, after a significant number of measurements, the state is strongly disturbed and each measurement is done over a noisy state and gives no information about the initial state.
Hence, there is an optimal number of measurements $t^{\mathrm {max}}(N,\Delta)$ which gives a maximal score $G^{\mathrm {max}}(N,\Delta)$. 
Moreover, for a fixed $N=2J$, $G^{\mathrm {max}}(N,\Delta)$ increases smoothly  as the measurements are weaker, i.e. as $\Delta$ increases. It reaches a limit $G^{\mathrm {max}}(N)$ (see Fig.~\ref{Goptsaturation} (a)) This suggest that for weak enough measurements, we observe the same behavior as in the 1D case. More measurements compensate a weaker interaction strength, without loss of precision. Hence, the precision of a single measurement is not important, as long as the measurement is weak enough. Moreover, in that case, we observe a plateau which suggests that the exact value of $t$ is not important. For $N\gg1$, even with $t$ far from $t^\mathrm {max}$, the mean score is close to $G^{\mathrm {max}}$.
Interestingly, the trade-off between $t^{\mathrm {max}}$ and $\Delta$ found for the 1D case seems to repeat here. We numerically find that $\sqrt{t^{\mathrm {max}}}/\Delta$ is constant for a given $N=2J$ (see Fig.~\ref{Goptsaturation} (c)) and scales as $1/\sqrt{N}$.

\begin{figure}
\center
\includegraphics[scale=0.7]{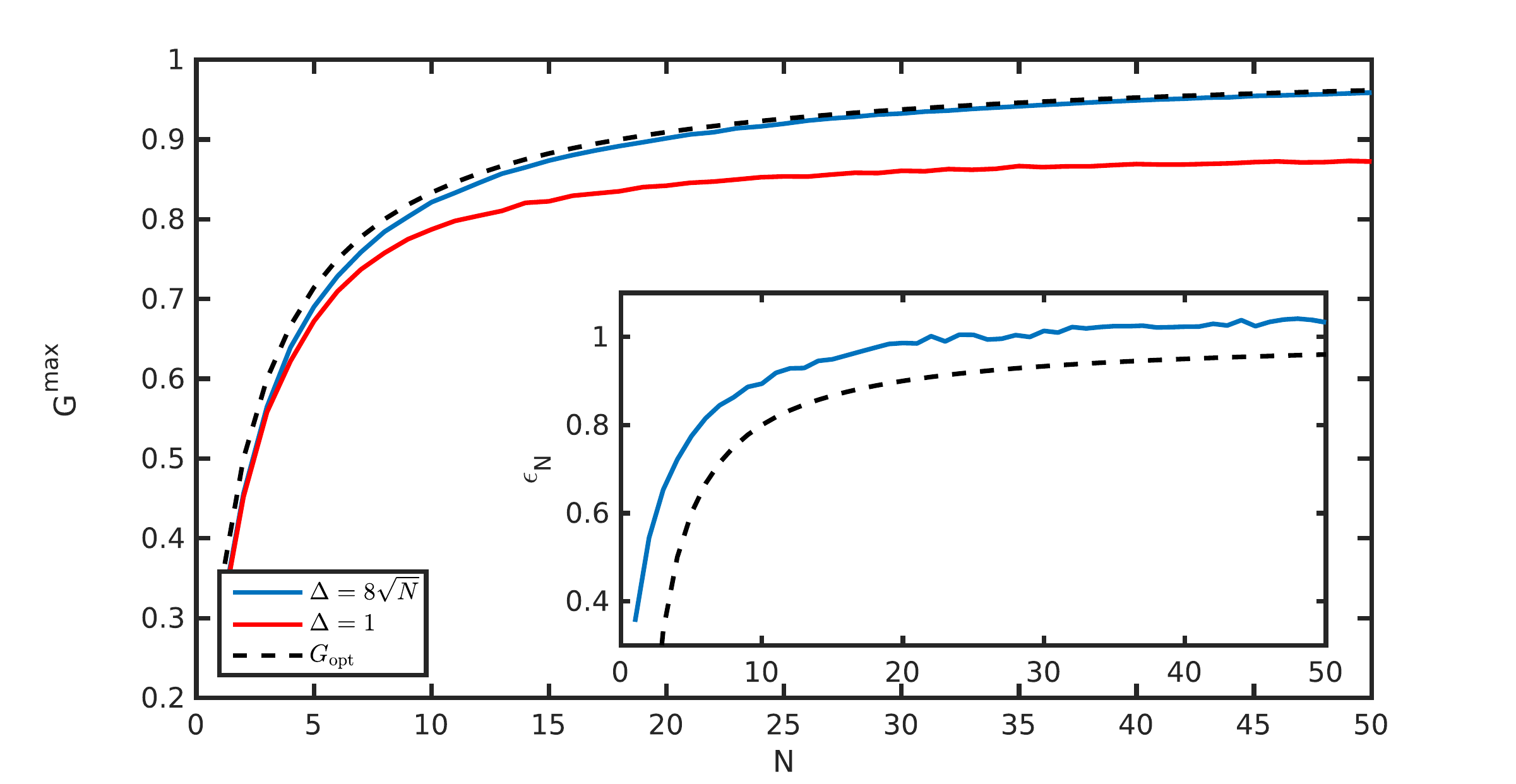}
\caption{Mean score $G^{\mathrm{max}}$ as a function of $N$ maximized over $t$. A too strong measurement ($\Delta=1$) fail to achieve an optimum. A weak enough measurement ($\Delta=8\sqrt{N}$) achieves a good score. The insert show $\epsilon_N= \frac{N}{2}(1-G_N)$. See Sec.~\ref{ultraweak3D} for further details.}
\label{Consecutive3dVariousDelta}
\end{figure}

Most importantly, for weak enough measurements, the obtained score is close to the optimal one, as shown in Fig.~\ref{Consecutive3dVariousDelta}. Numerical fluctuations prevent any precise statements about an estimation of the error, but the error is close to what was obtained with the initial measurement procedure, see Fig.~\ref{Consecutive3dVariousDelta}.

\section{Conclusion}

In this paper, we asked the question of how to model everyday measurements of macroscopic system within quantum mechanics. We introduced the notion of Macroscopic Quantum Measurement and argued that such a measurement should be highly non invasive, collect a large amount of information in a single shot and be described by a "fairly simple" coupling between system and observer. 
We proposed a concrete model based on a pointer von Neumann measurement inspired by the Arthur-Kelly model, where a pointer is coupled to the macroscopic quantum system through a Hamiltonian and then measured. This approach applies to many situations, as long as a natural Hamiltonian for the measured system can be found.

Here, we focused on the problem of a direction estimation. The Hamiltonian naturally couples the spin of the macroscopic quantum state to the position of a pointer in three dimensions, which is then measured. This reveals information about the initial direction of the state. We extended our previous study to consider a collection of aligned spins, which exploits the non monotonic behavior of the mean score as a function of the coupling strength. We presented more precise results. We relaxed the assumptions about the measured system, by considering a thermal state of finite temperature, and showed that our initial conclusions are still valid. We also relaxed the assumptions over the measurement scheme, looking at its approximation by a repetition of ultra weak measurements in several orthogonal directions. Here again, we obtained numerical results supporting the initial conclusion.
In summary, this MQM proposal tolerates several relaxations regarding lack of control or knowledge.

It is likely that these two relaxations can be unified: polarization measurement of systems with unknown number of particle or temperature should be accessible via the repeated 1D ultra-weak measurement way. However, this claim has to be justify numerically. Further open questions include the behavior of Arthur Kelly models in other situation where two or more noncommuting quantities have to be estimated, e.g. for position and velocity estimation.

\section*{Acknowledgements}

We would like to thank Tomer Barnea for fruitful discussions.
Partial financial support by ERC-AG MEC and Swiss NSF is gratefully acknowledge.

\section*{Author Contributions}
Nicolas Gisin suggested the study. Marc-Olivier Renou and Florian Fröwis performed
the simulations and worked out the theory. Marc-Olivier Renou wrote the paper. All authors discussed the
results and implications and commented on the manuscript at all stages. All authors have read and approved the
final manuscript.

\bibliography{main}

\end{document}